\documentclass[aps,prd,preprint,superscriptaddress]{revtex4-1}
\usepackage[dvips]{graphicx}
\usepackage{amssymb}
\usepackage{multirow}
\usepackage{longtable,amsmath}
\usepackage{textcomp}
\usepackage{xcolor}
\usepackage{soul}
\usepackage{todonotes}
\usepackage{setspace}
\usepackage{hyperref}

\begin{document}

%\preprint{}

\title{Pioneer 10 and 11 orbit determination analysis shows no discrepancy with Newton-Einstein's laws of gravity}

% Please use the appropriate macro for each type of information

\author{D. Modenini and P. Tortora}
\email[]{dario.modenini@unibo.it, paolo.tortora@unibo.it}
%\thanks{}
\affiliation{Department of Industrial Engineering, University of Bologna, Via Fontanelle 40, Forl\`i, Italy}


\date{\today}

\begin{abstract}
The present work describes the investigation of the navigation anomaly of Pioneer 10 and 11 probes which became known as the “Pioneer Anomaly”. It appeared as a linear drift in the Doppler data received by the spacecraft, which has been ascribed to an approximately constant sunward acceleration of about $8.5 \times 10^{-13} km/s^2$. %~\cite{Anderson2002}.
Since then, the existence of the anomaly has been confirmed independently by several groups and a large effort was devoted to find its origin. 
The present study consists of two main parts: thermal modeling of the spacecraft throughout its trajectory, and orbit determination analysis. Based on existing documentation and published telemetry data we built a thermal finite element model of the spacecraft, whose complexity has been constrained to a degree allowing for sensitivity analysis, leading to the computation of its formal uncertainty. The trajectory analysis and orbit determination was carried out using NASA/JPL's ODP (Orbit Determination Program) and our results show that orbital solutions may be achieved that do not require the addition of any "unknown" acceleration other than the one of thermal origin.

\end{abstract}

%\keywords{}

\maketitle

\section{Introduction}
The so-called Pioneer Anomaly is an anomalous blue shift in Pioneer 10 and 11 radiometric tracking data which has been interpreted as a constant sunward acceleration pulling the probes back during their journey towards and beyond the bounds of the Solar System. 
Since its discovery~\cite{Anderson1998}, the existence of the anomaly has been confirmed independently by several authors, see e.g. ~\cite{Markwardt2002},~\cite{Toth2009a},~\cite{Levy2009}. Since no conventional effects (e.g. unaccounted on-board or environmental systematic effects) were found to be completely satisfactory some authors have been suggesting more unconventional causes: these include modifications of the gravity law at scales of the Solar System size, or even the presence of dark matter, see for example~\cite{Reynaud2008} or~\cite{Brownstein2006}. The Pioneer Anomaly is considered by many authors as a deviation from Newton-Einstein's gravity law.

Here we show that no anomalous acceleration acted on the spacecraft and its evidence reported in many papers, is due to a lack in modeling Pioneer spacecraft dynamics: in particular, a model of recoil force due to anisotropic thermal radiation emitted must be added.

In last years several authors suggested as a possible on-board effect for explaining the anomaly the recoil force due to anisotropic infrared emission. Even if in their early investigation ~\cite{Anderson1998} Anderson et al. discarded the thermal recoil force (TRF) as a source of a significant bias acceleration, its impact was reconsidered soon after by Katz in his comment ~\cite{Katz1998}. The author stated that the recoil due to the fraction of thermal power radiated by the radioisotope thermoelectric generators (RTGs) being scattered from the back of the spacecraft antenna, together with the on-board dissipated electrical power radiated from the back of the spacecraft were compatible with the reported anomalous acceleration. This conclusion, however, was disputed by Anderson et al. in their response ~\cite{Anderson1999}. While both the above contributions were mainly semi-quantitative, only more recently the study of the TRF acting on Pioneer spacecraft was the subject of deeper analyses, such as the ones in ~\cite{Scheffer2003},~\cite{Bertolami2010},~\cite{Rievers2010},~\cite{Toth2009}. On one side, some authors provided estimate of recoil force and compared it with the reported magnitude of the anomaly to check which part of the anomalous acceleration could be explained by recoil force. On the other side (~\cite{Turyshev2011}) started from the observation that, if a force of thermal origin actually acted on the probes, it should have exhibited a time decrease following the nuclear decay of plutonium (the source of power for the probes), to show that the observed drift in Doppler residuals is compatible with accelerations varying on that time scale.
A recent paper by Turyshev et al. (\cite{Turyshev2012}) provides a consistent combination of the previous two approaches, by making use of an extremely sophisticated thermal model of the spacecraft to be integrated within the tracking data analysis and orbit determination process as an additional dynamical model. Their conclusion is that the thermal recoil force due to the  anisotropic infrared emission is the cause of the drift of the Doppler residuals, which gave rise to the so-called "Pioneer Anomaly".

The work presented herein is an independent and parallel analysis with respect to the one carried out by Turyshev et al., still fully consistent with their approach. However, our analysis goes beyond and deepens the results obtained in \cite{Turyshev2012} in two respects: 
\begin{itemize}
\item an independent spacecraft thermal model was built by using a finite element model of the probes, whose complexity was constrained to a degree allowing for sensitivity analysis and computation of uncertainties;
\item analysis of the most complete data sets of both Pioneer 10 and Pioneer 11 radiometric Doppler data in an orbit determination process where the thermal recoil force resulting from our thermal model is included (along with its uncertainty) as an additional dynamical model; this latter effort is in particular finalized to answer the question whether any additional empirical acceleration would still be needed to obtain zero-drift Doppler frequency residuals.
\end{itemize}
The paper is organized as follows. Section \ref{sec:thermal model} discusses the thermal model of Pioneers spacecraft, presenting the fundamental theoretical aspects and the modeling steps; focus is placed on the aspects related to the integration of the results into the ODP. In Section \ref{sec:Orbital} the basics of the orbit determination theory are covered; the implementation details (filtering techniques, data editing) are deeply discussed. In \ref{sec:Results} the results of the analyses of Pioneer 10 and 11 Doppler tracking data are presented and examined. Finally, in \ref{sec:Conclusions}, conclusions drawn from the overall investigation are discussed and summarized. 

\section{Pioneers thermal model}
\label{sec:thermal model}
\begin{figure}[h!]
  \caption{Prototype of Pioneer 10 hanged in the Smithsonian National Air and Space Museum}
   \label{fig:Pio1}
    \includegraphics[width=0.75\textwidth]{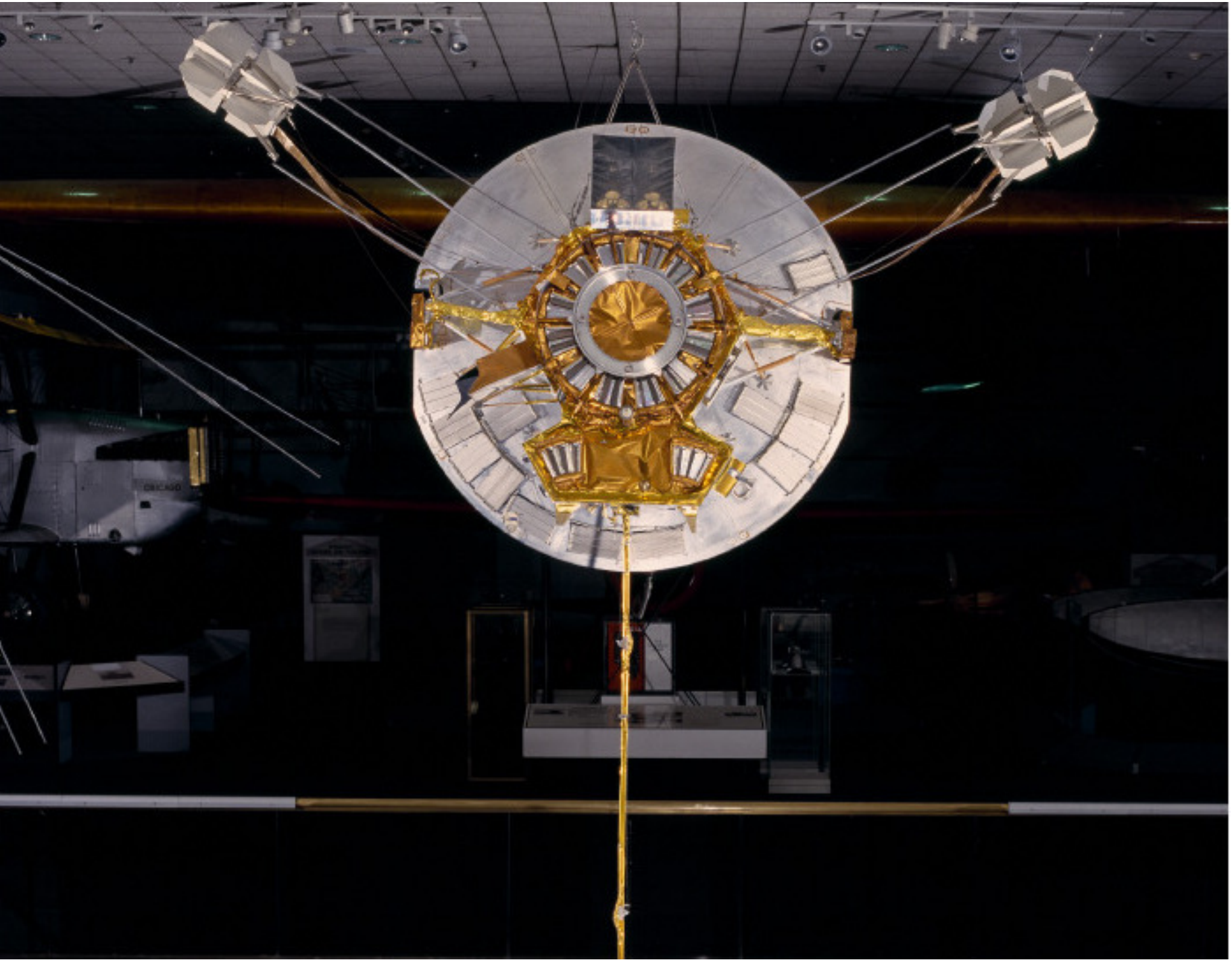}
\end{figure}

\begin{figure}[h!]
  \caption{Pioneer F/G spacecraft main components from ~\cite{PC-202}}
  \label{fig:Pio2}
    \includegraphics[width=0.75\textwidth]{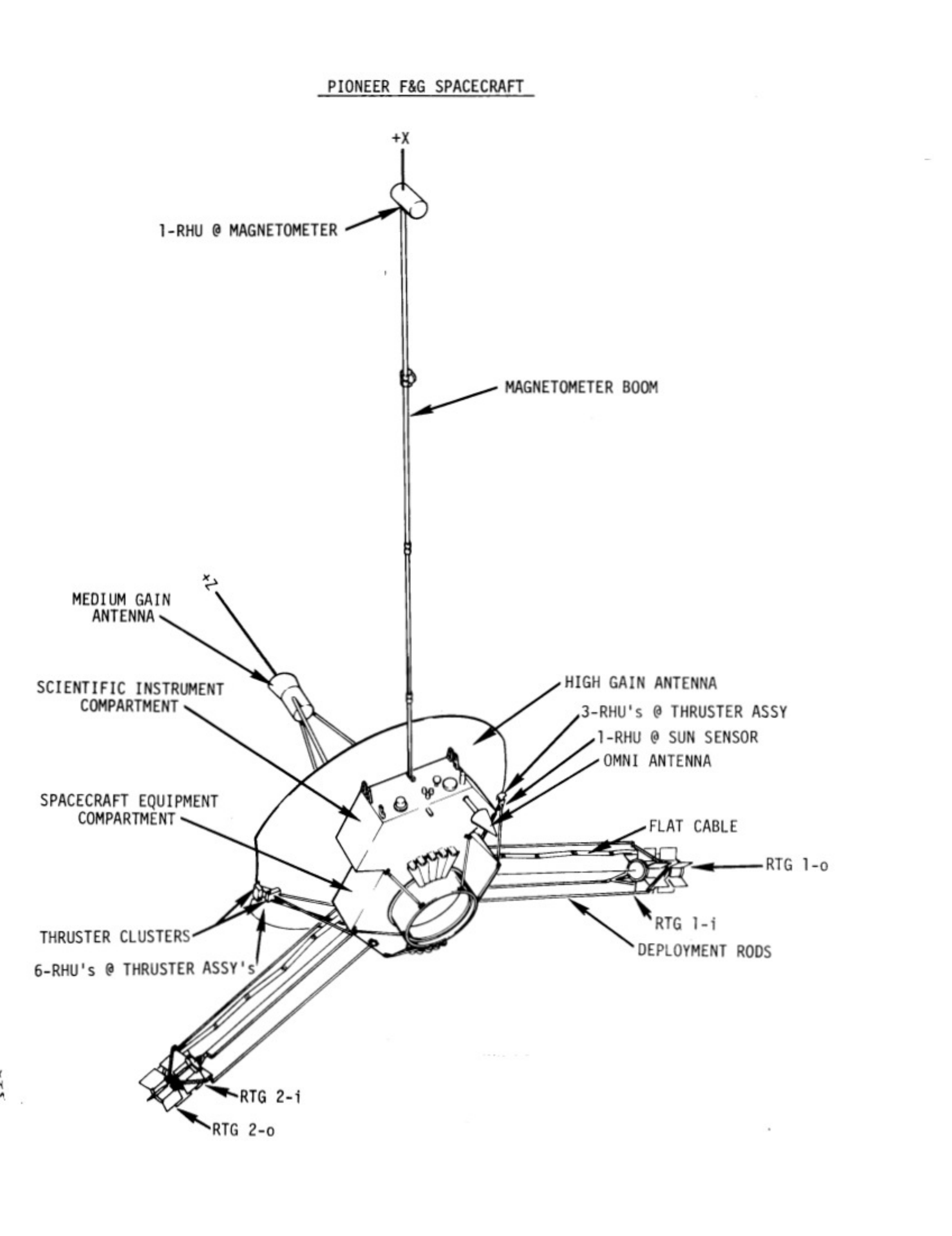}
\end{figure}

Pioneer spacecraft are depicted in FIG. ~\ref{fig:Pio1} and ~\ref{fig:Pio2}. The main sources of power on board Pioneer spacecraft, are four radioisotope thermoelectric generators. The thermal power generated by RTGs was nearly 2580 W at launch; its amount is expected to decay during the mission following plutonium half-life time (87.7 years). A fraction of this thermal power ($\approx$ 160 W at launch) is converted into electrical power which supplies the various instrumentation placed on board. In particular, part of it is transmitted towards the Earth as radio beam, while the remaining part is converted into heat by Joule effect: this waste heat is radiated into space through the main compartment external surfaces and a louver system which ensures thermal control of the spacecraft. In other words, all power generated inside the Pioneers spacecraft is expelled in form of electromagnetic radiation (either IR or radio beam), which carries momentum with it. If the radiation pattern is anisotropic, the momentum exchange between the spacecraft and the emitted radiation results in a recoil force which affects the trajectory of the probe.


The component of the recoil force on a certain direction equals the unbalanced power output in the same direction divided by the speed of light. Therefore a simple order of magnitude analysis tells us that, the mass of the spacecraft being $\approx 250 Kg$, only a small fraction of the total available power, $\simeq 60 W$, directionally radiated away from the sun would cause an acceleration equal to the anomaly. It is hence clear that recoil force due to radiation should be properly estimated and included in the orbit determination process.

By inspection of the macroscopic configuration of Pioneer geometry as visible in FIG. ~\ref{fig:Pio1} and ~\ref{fig:Pio2} one can identify at least two mechanisms which are likely to be responsible of a certain degree of anisotropy in radiation emission:
\begin{enumerate}
 \item Heat from RTGs rejected by the highly reflective backside of the high gain antenna;
 \item Electrical heat dissipated inside the bus having a preferable escape direction through the louver system.
\end{enumerate}
Both these contributions are mostly directed as the anomaly, but while the first depends on thermal power, the second depends on electrical power. There is actually another source of thermal energy on-board other than the RTG’s which are 9 radioisotopes heaters units (RHU) generating $1W$ each, and deputed to heat up thruster cluster assemblies. However, according to Turyshev ~\cite{Turyshev2010} their geometric configuration and location is such to prevent them to contribute substantially to anisotropic radiation. 
In earlier papers ~\cite{Anderson2002} it was pointed out that the secular time evolution of RTG’s power is quite in contrast with the evidence of a constant anomaly, even if Markwardt's analysis ~\cite{Markwardt2002} of the set of data between 1987 and 1998 claims that the Doppler shift is compatible with a jerk term of a time scale similar to the decay of Plutonium. More recently, Turyshev ~\cite{Turyshev2011} arrived to a similar conclusion using extended data sets of both Pioneer 10 and 11 and testing for a linear and an exponential time decay of the anomaly. 
In the present work, we also explored the temporal variation of the recoil force of thermal origin; however, we did not restrict to the case of a monotonic decrease, as could be expected for an effect driven only by the available on-board power. The spacecraft infrared emission depends on the total energy input to the spacecraft which, besides RTG's power, also includes irradiation from the Sun. The solar flux effect on spacecraft dynamics is twofold: on one side there is the solar radiation pressure, which is commonly accounted for during trajectory integration. The ODP implements a model to estimate the momentum exchange between the solar flux and the spacecraft components. On the other hand the solar flux induces a temperature increase on the illuminated surfaces due to the fraction of radiation which is adsorbed. For the Pioneers, this is the case especially for the high gain antenna which is constantly pointed towards Earth and also fully illuminated by the Sun, at least at sufficiently large heliocentric distances. The parabolic dish is basically a thin surface with highly different emittance on the two sides: the back side of the antenna is highly reflective with a low emittance, while the white painted front-side is highly emittive (see Table ~\ref{tab:Table1}). Therefore the solar power is almost entirely dissipated by the Earth-pointing antenna face, resulting in a recoil force anti-parallel to the contribution of the internally generated power, hence subtracting from its amount. The subtracting term due to the solar flux decreases with time as well because of the probes receding from the Sun, such that, as a result, our thermal model reconstructs a recoil force which does not decrease with time monotonically.

%The thermal model of the spacecraft is built based on a FEM software for heat transfer analyses. 
\subsection{Thermal radiation theory for spacecraft modeling}
The thermal model of the spacecraft includes a simplified discretized geometry of the probe, which has 
been developed based on the existing design documentation. The thermal state has been reconstructed from published recovered telemetry data ~\cite{Turyshev2010} which consist of measurements of power generation/consumption and temperature readings from several sensors located on the RTG’s fins, the body panels and inside the payload bay. However, while the energy input has been used quantitatively, the temperature information has been used only for a qualitative comparison with the thermal model predictions.

To determine the amount of radiation emitted by the spacecraft into space, the exchange of thermal energy between the different surfaces is required. The mathematical details behind this computation are of no interest here and only the relevant modelling steps are presented. The process involves computation of the radiation balance on each element between emitted, adsorbed and reflected radiation, and of the so-called view factors between the discretized surface elements. The former can be computed following the well-known Stephan-Boltzmann law: 
\begin{equation}
\label{Steph-Boltz}
 q_{rad}=\sigma \varepsilon T^4
\end{equation}
Where $\sigma$ is the Stephan-Boltzmann constant equal to $5.67\cdot10^{-8} W/(m^2K^4)$, $\varepsilon$ is the surface total hemispherical \footnote{Averaged over all directions and wavelengths of radiation} emittance, and $T$ is the absolute temperature.
When radiation impinges over an opaque surface, Eq. \eqref{Steph-Boltz} should be modified to account for the reflected and absorbed radiation such that the radiation energy balance can be expressed as follows:
\small
\begin{equation}
\label{RadBal1}
 q_{rad} = J-H = (1-\varepsilon)H + \varepsilon \sigma T^4-H= \varepsilon(\sigma T^4-H)
\end{equation}
\normalsize
In Eq. \eqref{RadBal1} $J$ denotes the surface radiosity, i.e. the total heat flux leaving the surface due to emission and reflection, and with $H$ the surface irradiation, that is the total incoming heat flux~\cite{ModestBook}; moreover, the assumptions of gray and diffuse emitting/reflecting surfaces have been retained (so called lambertian radiators \footnote{A lambertian emitter is a surface for which the directional radiative flux (energy flow per unit angle per unit surface area), is proportional to the cosine of the polar angle ~\cite{ModestBook}. In an analogous manner can be defined a lambertian reflector. Gray assumption further implies that reflectance $\rho$ equals $1 - \varepsilon$.}).
The view factors provide a measure of the amount of the total radiation which, emitted by a surface, hits another surface after mutual shadowing. To compute them one needs to know, other than the relative orientation between the surfaces, also the directional dependency of the emitted radiation. 
Under these hypotheses, the radiation pressure acting on an isolated flat surface element of area $dA$ due to emitted radiation is given by the following expression:
\begin{equation}
 p_{rad}=\frac{2}{3c} \sigma \varepsilon T^4
\end{equation}
Where $c$ is the speed of light. The force due to such pressure acts in the surface normal direction.

Radiation pressure can be generalized to account for absorption and reflection, such that a radiation recoil force can be computed at each surface element to be then integrated over the entire spacecraft surface. An alternative approach used in the present work, consists of surrounding the spacecraft with a sphere acting as a control volume. This control volume is modeled as a “passive” blackbody, that is a body at a constant temperature of 0 K and emittance of 1, such that it absorbs all the incident radiation without emitting or reflecting anything: the net radiation escaping out of the spacecraft system and detected by the control volume is the only contribution to the recoil force since the contribution due to radiation intercepted (absorbed) by spacecraft components cancels out the pressure acting on the surfaces emitting such radiation. Conservation of energy requires that volumetric ($ Q $) and surface ($\vec{q}$) heat sources input to the the spacecraft balance the net radiation emitted by the spacecraft itself and impinging on the control volume (CV):
\small
\begin{equation}
\label{RadBal2}
 \iiint \limits_{V_{Pio}} Q dV + \iint \limits_{S_{Pio}} \vec{q}_{Sun} \cdot \vec{n} dS =  \iint \limits_{S_{Pio}} (J-H) dS = \iint \limits_{CV} H dS
\end{equation}
\normalsize
This equation has been used to check the consistency of the implemented thermal model.

\subsection{Spacecraft numerical model}
The geometric model created for the Pioneers includes only the major spacecraft components, namely the high gain antenna, the RTGs, the adaptor launch ring (ALR) and the spacecraft compartment bus plus the louvers radiators. The energy input to the system consists of three volumetric heat sources, two placed inside the RTG’s and one placed inside the spacecraft body, plus a surface heat flux to mimic the solar radiation impinging on the concave side of the high gain antenna and on the RTG’s (this last contribution is however negligible when compared to the thermal power inside RTG’s) which are the only parts actually exposed to it during the interplanetary cruise. As a baseline case study, we have assumed that, being known the total amount of power from telemetry and the volume of the components where the heat sources are placed, the distribution of such sources is uniform within each component. This is of course a simplification, especially for the electrical power inside the bus, since the presence of several instrumentation components makes the produced heat more likely being concentrated in some regions. The temperature readings from the six sensors placed on the bus platform indicates temperature differences up to 30 \textcelsius\ among the different locations. The impact of such temperature differences on the computed radiation pattern has been addressed by performing a set of Monte Carlo simulations, resulting in a relatively limited scatter of the recoil force, as it will be detailed in Section \ref{sec:MC}.

The thermo-optical properties of surface materials were retrieved mainly from ~\cite{PC-202}, and the relevant ones are reported in Table \ref{tab:Table1}. There are some minor differences between the values reported in ~\cite{PC-202} and what actually used in the present study, and these are reported between round brackets. In particular, the nominal values of solar absorptance are beginning-of-life values, which are likely to vary during cruise as an effect of surface degradation due to exposure to UV radiation, charged or contaminating particles. The generalized result is an increase in the solar absorptance \footnote{usually the effect on the emittance is small, unless severe deposition of dust occurs such to sensibly modify the surface coating (~\cite{thermal_handbook})}. In ~\cite{thermal_handbook} white paint absorptance is reported to increase from $\approx 0.20$ up to $\approx 0.60$ in few years of mission. Moreover, for the Multi Layer Insulation (MLI), the design documentation reports the emittance of the external layer, 0.70, while, as pointed out in ~\cite{Scheffer2003}, an effective emittance should be used instead which lies in the range 0.007 $\div$ 0.01. 
 \begin{table*}%[H] add [H] placement to break table across pages
 \caption{Thermo-optical properties of main Pioneers' surfaces relevant to thermal model}
 \label{tab:Table1}
 %\begin{small}
 \begin{ruledtabular}
 \begin{tabular}{ l  c  l  c  c}
 %Component & Material &  Surface coating & $\epsilon$ & $\alpha$ \\
 Element & Material &  Surface coating & $\epsilon$ & $\alpha$ \\
 \hline  
 HGA Front side & Al 6061  &    DC92-007 white paint &  0.85  &  0.21 (0.50) \\
 HGA Back side &  Al 6061  &      white paint         &  0.04  &  0.17 \\
 RTG body &  HM31A-F Mg &  white paint 		& 0.82  &  0.21 (0.50) \\
 RTG fins &  HM21A-T8 Mg &  white paint 	& 0.82  &  0.20 (0.50) \\
 S/C MLI &  Al 6061 	& aluminized Mylar/Kapton & 0.69 (0.0085)  &  0.20/0.46 \\
 Louvers &  Al 6061   &  bare 			&   variable & -  \\        
 ALR interior  & Al 6061 & black paint & 0.84   &  0.95 \\
 ALR exterior    &  Al 6061 &  bare  	 &  0.10  &  0.24 
% Lines of table here ending with \\
 \end{tabular}
 \end{ruledtabular}
 %\end{small}
 \end{table*}
 
\begin{figure}[h!]
  \caption{Heat dissipated by Louvers, taken from ~\cite{Turyshev2010}}
  \label{fig:fig3}
    \includegraphics[width=0.6\textwidth]{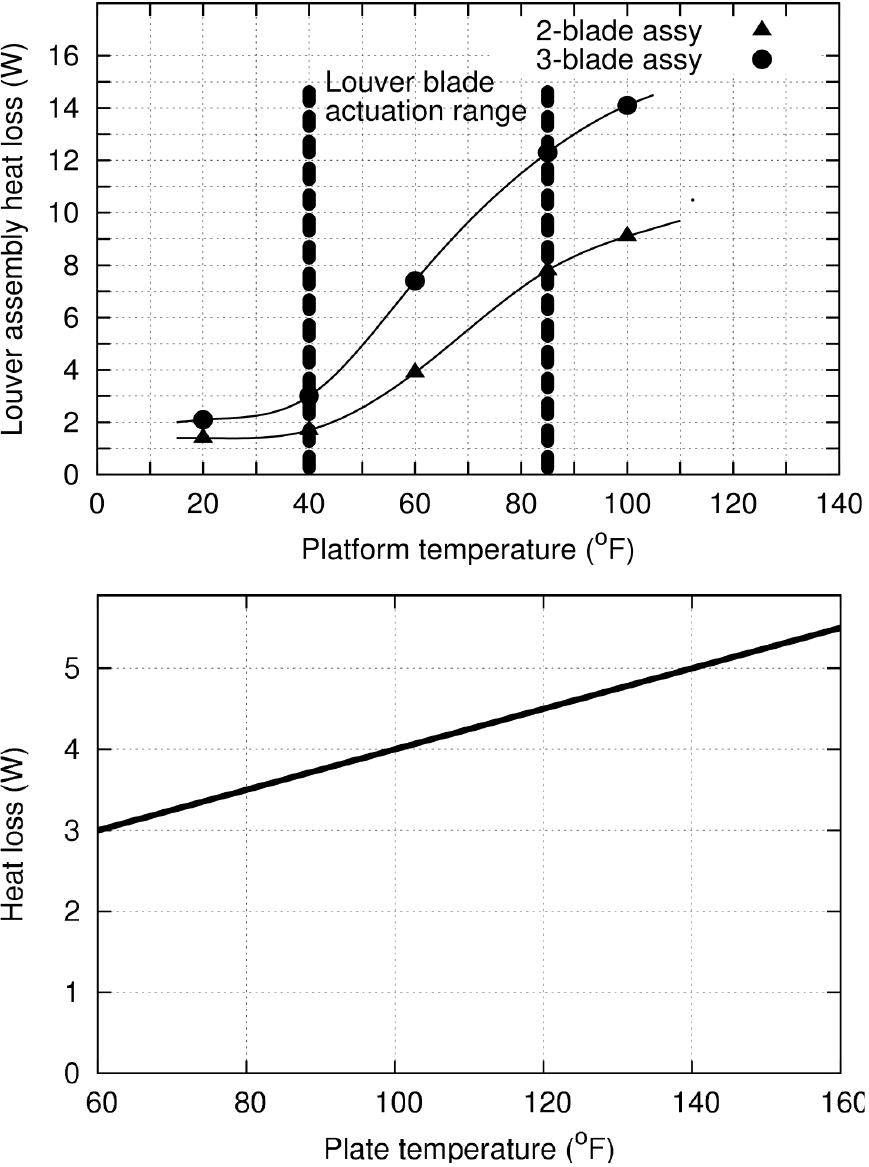}
\end{figure}

The presence of louvers system has been simulated by specifying a variable emittance over a region surrounding the ALR, which is a function of the temperature and spatial coordinates. Therefore, no detailed geometric components for blades, springs and platform have been used. It has been
rather preferred to use as driving information the amount of Watts dissipated by the louver system, as a whole, across its operating temperature range (from 4 to 32 \textcelsius). Such quantities can be retrieved from the plots in FIG. ~\ref{fig:fig3}, which were taken from ~\cite{Turyshev2010}: the first reproduces the power radiated by each 2-blade and 3-blade assembly, while the second provides heat loss from the louvers structure. Based on these diagrams, the following empirical function for emittance variation was implemented in the thermal model:

\small
\begin{equation}
\label{eq2}
\varepsilon = 0.01+0.58- \frac{0.38}{1+ \exp (0.35(T-288))} e^{-10|x^2+z^2-0.4^2|}
\end{equation}
\normalsize

The function used in Eq. \eqref{eq2}, S-shaped exponential for temperature, Gaussian bell-shaped for spatial coordinates, was arbitrarily selected just to avoid sharp discontinuities. Outside the temperature actuation range, the emittance is kept constant, which seems a reasonable assumption. The numerical coefficients in \eqref{eq2} were tuned in order to match the value of the radiated power extrapolated from the top panel in FIG. ~\ref{fig:fig3}, at four measured temperatures. From Stephan-Boltzmann law follows that prescribing the radiated power at a certain temperature is equivalent to prescribe the area integral of the emittance, $\int \varepsilon dA$: the computed values are collected in Table \ref{tab:Table2}.

\begin{table}%[\nsuccH] add [H] placement to break table across pages
 \caption{Heat loss from louvers and related emittance surface integrals}
 \label{tab:Table2}
% \begin{ruledtabular}
 \begin{tabular}{c c r}
 \hline
 \hline
  T platform  & Louvers    & $ \int \varepsilon dA $ \\
    $[K]$     & heat loss $[W]$  &  $[m^2]$ \\
 \hline
  266 &  20 & 0.07 \\
  278 & 30 & 0.09 \\
  288.7 &  64 & 0.16 \\
  303 & 124 & 0.26 \\
  \hline
  \hline
 \end{tabular}
%\end{ruledtabular}
\end{table}%[H] add [H] placement to break table across pages

Equation \eqref{eq2} indicates that the louvers equivalent emittance equals 0.21 in fully closed configuration and 0.59 in fully open condition. These numbers are somehow in disagreement with data found in ~\cite{PC-202} itself, which reports $\varepsilonε = 0.04$ for the blades, and 0.82 for the radiating platform underneath. In this respect, we explicitly note that the surface integral of Eq. \eqref{eq2} is highly dependent on the type of mesh used in the numerical model; in other words, the relevant physical data is the value of $\int \varepsilon dA,$ while the resulting emittance is intimately related to, and a consequence of, the actual discretized geometry. The inconsistency is therefore only apparent. Once again it is noted that the driving criterion is the preservation of the total power emitted by the louvers according to the design documentation, rather than matching the detailed geometry, surface area or emittance separately. It is certainly true that the radiated power can be indirectly inferred by temperature, area and emittance of a surface, but, on the other hand, when a direct measure of the radiated power is available, as for the louvers, it seems reasonable to prefer this source of information.

The output of the full thermal model is the amount of radiation which, after mutual reflection among surfaces, escapes the spacecraft system to hit the control volume. The integral of the radiation flux impinging the control volume and projected along the high gain antenna axis direction, provides the magnitude of the anisotropic emitted power.

\subsection{Thermal simulations output sensitivity and covariance analyses}
\label{sec:MC}
To assess the sensitivity of the thermal model to parameters which are not exactly known, and as a method to provide confidence bounds for the results, a series of Monte Carlo simulations and a covariance analysis were set up. The objective of this process is to obtain a fit of the recoil force as a function of thermal power, $P_{th}$, electrical power, $P_{el}$, and solar flux $\Phi_S$ (known input parameters) while a number of other parameters are allowed to vary over the simulations over a certain space (uncertain parameter space). These include surface emittance and absorptance, and power distribution within volumes. 
In particular, allowing for a non-uniform power distribution inside a component is a way to mimic, within the frame of our thermal model, the presence of a spatial variation in the temperature over the bus and the RTG's, which was indeed highlighted by the Pioneers telemetry temperature data.
The volumetric heat sources distribution functions inside the RTG's ($Q_{th}$) and the spacecraft bus ($Q_{el}$) \footnote{$Q_{el}$ and $Q_{th}$ are volumetric heat coefficients related to the total thermal and electrical powers as follows:
\small
\begin{equation*}
P_{th}=\iiint_{RTGs} Q_{th}dV_R, \quad P_{el}=\iiint_{Bus} Q_{el}dV_B
\end{equation*} } 
\normalsize
were arbitrary assumed as being sinusoids of the spatial coordinates. The amplitude and phases of such sinusoids belong to the uncertain parameters space \cite{Modenini2012}. 

The solar absorptances of the high gain antenna and of RTG’s were varied according to:

\begin{equation}
%\label{eq3}
\alpha_i = \overline{\alpha}_i + d\alpha_i
\end{equation}

Where $\overline{\alpha}_i$ is the nominal absorptance set equal to 0.5 and $d \alpha_S$ are sampled drawn from uniform distributions in the open interval (-0.1, 0.1).
The uncertainty in surfaces emittance coefficients was set to 15\% of the nominal value:

\begin{equation}
%\label{eq3}
\varepsilon = \overline{ \varepsilon }_i + d\varepsilon_i
\end{equation}
Where $\overline{ \varepsilon }$ is the nominal emittance of the i-th modeled surface and the values for $d\varepsilon_i$ are sampled from a uniform distribution inside the interval (-0.15$ \overline{\varepsilon_i} $, 0.15$ \overline{\varepsilon_i} $).

Numerical consistency of the method has been checked according to Eq. \eqref{RadBal2}. Our results indicate that while the first equality is always satisfied (difference below $10^{-7} W$), the second equality gives residuals in the order of 2 to 3 W, which reflects in about 0.1\% of the total input power. If one assumes this energy unbalance having no preferential direction, the discrepancy per unit solid angle is  $3W/4 \pi \simeq 0.23 W$. This value can be assumed as a figure of merit of the numerical error affecting the computed directional radiated power.\par
 
In order to be incorporated in the orbit determination process, the results of the Monte Carlo simulations need to be converted in an acceleration fitted by a function of time which can be represented within the ODP (i.e. polynomials up to the $4^{th}$ order and exponential functions). In such a way, the recoil acceleration is represented by a finite number of parameters and their associated covariance matrix. Since it is reasonable to expect the directionally radiated power, $W_Z=TRF\cdot c$, to be a linear function of the energy input to the system, one can then seek for a regression of the Monte Carlo simulations output of the kind:

\begin{equation}
\label{TRF_1}
W_Z = \pmb{x} \cdot \pmb{P} = x_1P_{th}+x_2P_{el}+x_3\widetilde{\Phi}_S
\end{equation}
where $x_i, i=1,2,3$ are coefficients to be determined. To preserve homogeneous dimensions, a solar power, $\widetilde{\Phi}_S$, has been introduced,  equal to the solar flux times the projected surface of the HGA. This approach is similar to what used by Turyshev et al. in \cite{Turyshev2012}, except that here we explicitly account for the solar power contribution in the fit of the thermal simulation. 

A least-squares fit of the Monte Carlo results provided the vector of regression coefficients $\pmb{x}$ with associated covariance matrix $\varGamma_x$, as shown in Table \ref{tab:Table3}.

\begin{table}%[H] add [H] placement to break table across pages
 \caption{Regression coefficients for anisotropic IR emission of Pioneers probes as a function of the on board power and solar flux}
 \label{tab:Table3}
 %\begin{ruledtabular}
 \begin{tabular}{ c  c  c  }
 \hline
 \hline 
  Estimated $\pmb{x}$ & Estimated $\pmb{\sigma}$  & Normalized covariance $\varGamma_x$ \\
  \hline
 % \hline 
 0.0132 &  $1.76 \times 10^{-4}$ & \multirow{3}{*}
{$ \begin{bmatrix} 
  1 & -0.905 & 0.195 \\
  -0.905& 1 &-0.478 \\ 
  0.195 & -0.478 & 1 \\
\end{bmatrix} $} \\ 
 0.553 & $8.17 \times 10^{-4}$ \\
-0.207 &  $9.02 \times 10^{-3}$ \\
 \hline
 \hline
 \end{tabular}
%\end{ruledtabular}
\end{table}%[H] add [H] placement to break table across pages

A total of $n_{sim}=1000$ simulations were performed, each using a triplet ($P_{th}, P_{el}, \Phi_S $). 
The termination criterion used was the invariance of the resulting statistics: the least square fit was performed incrementally over the simulations, i.e. using a number of points ranging from 1 to $n_{sim}$, and the difference between two subsequent fit output monitored. The percentage variations of the elements of the regression vector and covariance matrix stabilized within 0.3\% after around 500 simulations.
As a measure of the scatter of the simulation results, one can look at the post-fit residuals after linear regression, which exhibit a standard deviation of $\simeq 4.5 W$, while their mean is $\simeq 0.1 W$ (to be compared to the magnitude of the anisotropic power, in the order of a few tens of Watt).
The covariance matrix shown in Table \ref{tab:Table3} globally accounts for the uncertainties in internal distribution of thermal power inside the RTGs and the electrical power inside the spacecraft body, as well as uncertainties in surface optical properties and, finally, the goodness of the assumed linear fitting function for the thermal recoil force. There are, however, other sources of error which may affect the thermal model. First, while the power values are known inputs to the thermal model, their temporal evolution during the trajectory is not perfectly known. Indeed, 
the telemetry readings from which the thermal and electrical power values are retrieved, have limited resolutions: according to ~\cite{Turyshev2010}, ~\cite{Toth2009} the confidence in $P_{th}$, $P_{el}$ is limited to 2.1 and 1.8 W, respectively, at 1-sigma level.
Moreover, the solar flux is not a measured quantity being rather estimated through an approximate relation assuming a decay proportional to the heliocentric distance squared ($\Phi_S=1366/AU^2$ $W/m^2$). Its uncertainty was modeled in this study solely in terms of the flux constant at 1 AU, to which a 1-sigma of $\pm 4 W$ was assigned \footnote{The periodic solar cycle induces a variation on the irradiance measured at the Earth's upper atmosphere of about 2 W (http://glory.gsfc.nasa.gov/overview-tsi.html). This number has been conservatively doubled as to compensate for the inaccuracy of the assumed quadratic decay}. Inclusion of all these error sources can be accomplished applying the theory of linear estimation in the presence of consider parameters ~\cite{Moyer1971}. The mathematical details will be skipped for the sake of brevity, suffice to say that the overall result is an additional covariance to be added to the one in Table \ref{tab:Table3}.

To integrate the thermal recoil acceleration in the ODP in a consistent manner, one needs to map the representation found in Eq. \eqref{TRF_1} as a function of power sources, to the time domain.
In ~\cite{Toth2009} plots of thermal power inside RTG’s (total expected power minus the telemetered electrical power) and of the electrical power dissipated from instrumentation placed inside the spacecraft body are found. Computation of the solar flux variation over time requires the spacecraft heliocentric distance, which grows almost linearly in time during interplanetary cruise: the exact variation is retrieved from orbital solutions for Pioneer 10 and 11 and fitted with a suitable function of time. A combination of polynomials and exponential functions was found to provide a satisfactory fitting of the time evolution of power data, and they are functions natively incorporated in the ODP as acceleration models. 

We mentioned above that the IR emission is not the only kind of energy radiated into space by Pioneer, since there is also the power, nominally $8W$, carried by the collimated radio beam transmitted by the high gain antenna. This aspect is discussed in  ~\cite{Scheffer2003} where an efficiency of the conversion from power to linear momentum of 0.83 is computed: this value has been used in the present study as well. The resulting force pushes the probe away from the sun, thus it has to be subtracted to the thermal recoil acceleration just computed. This way, a global radiation recoil force (RF) expression\footnote{In the following, for the sake of simplicity, we will continue to use the expression ``thermal recoil force'' keeping in mind that the recoil due to the transmitted radio-beam power is also included.}, to be integrated in the orbit analysis discussed hereafter, was computed for Pioneer 10 and 11 as follows:
\small
\begin{multline} 
\label{RadAcc}
% = \frac{W_z}{c \cdot m_{P_{10}}} 
a_{_{RF_{P_{10}}}} = A_{RF} + B_{RF}\tau + C_{RF}\tau^2 + G_{1RF}e^{- \beta_1 \tau_1}+G_{2RF}e^{- \beta_2 \tau_2} \\
a_{_{RF_{P_{11}}}} = A_{RF}+B_{RF}\tau + C_{RF}\tau^2 + A_{\phi_i} + B_{\phi_i}{\tau_i} + \\ 
+ C_{\phi_i}{\tau_i}^2 + D_{\phi_i}{\tau_i}^3 + E_{\phi_i}{\tau_i}^4, \; \; i=1,2
\end{multline}
\normalsize

In the above equation the first three terms account for the on board power contribution to radiation force, while the remaining (exponentials for Pioneer 10, $4^{th}$ order polynomials for Pioneer 11) account for the solar flux; $\tau =$ seconds past launch; $\tau_1 =$ seconds past $1^{st}$ September 1977, up to $\tau_2$; $\tau_2 =$ seconds past $12^{th}$ September 1980. In the expression applicable to Pioneer 11, two $4^{th}$ order polynomials have been used for fitting the two segments of trajectory covered by tracking data, prior and after the Saturn encounter. For convenience, we can collect the coefficients of the recoil force in a vector $\pmb{\xi}_{P10}= [ \begin{matrix} A_{RF} & B_{RF} & C_{RF} & G_{1RF} & G_{2RF} \end{matrix} ]$ for Pioneer 10, and in an analogous $\pmb{\xi}_{P11}$ for Pioneer 11. %The reason why coefficients of the recoil force have been used rather than for the acceleration will become soon clearer. 

\begin{figure}
  \caption{Acceleration due to emitted radiation by Pioneer 10 (full line) and Pioneer 11 (dash-dot line) along the trajectory segments covered by the analyzed tracking data. The gap in Pioneer 11 curve around 1979 corresponds to the Saturn encounter.}   \label{fig:TRAfig}
    \includegraphics[width=0.75\textwidth]{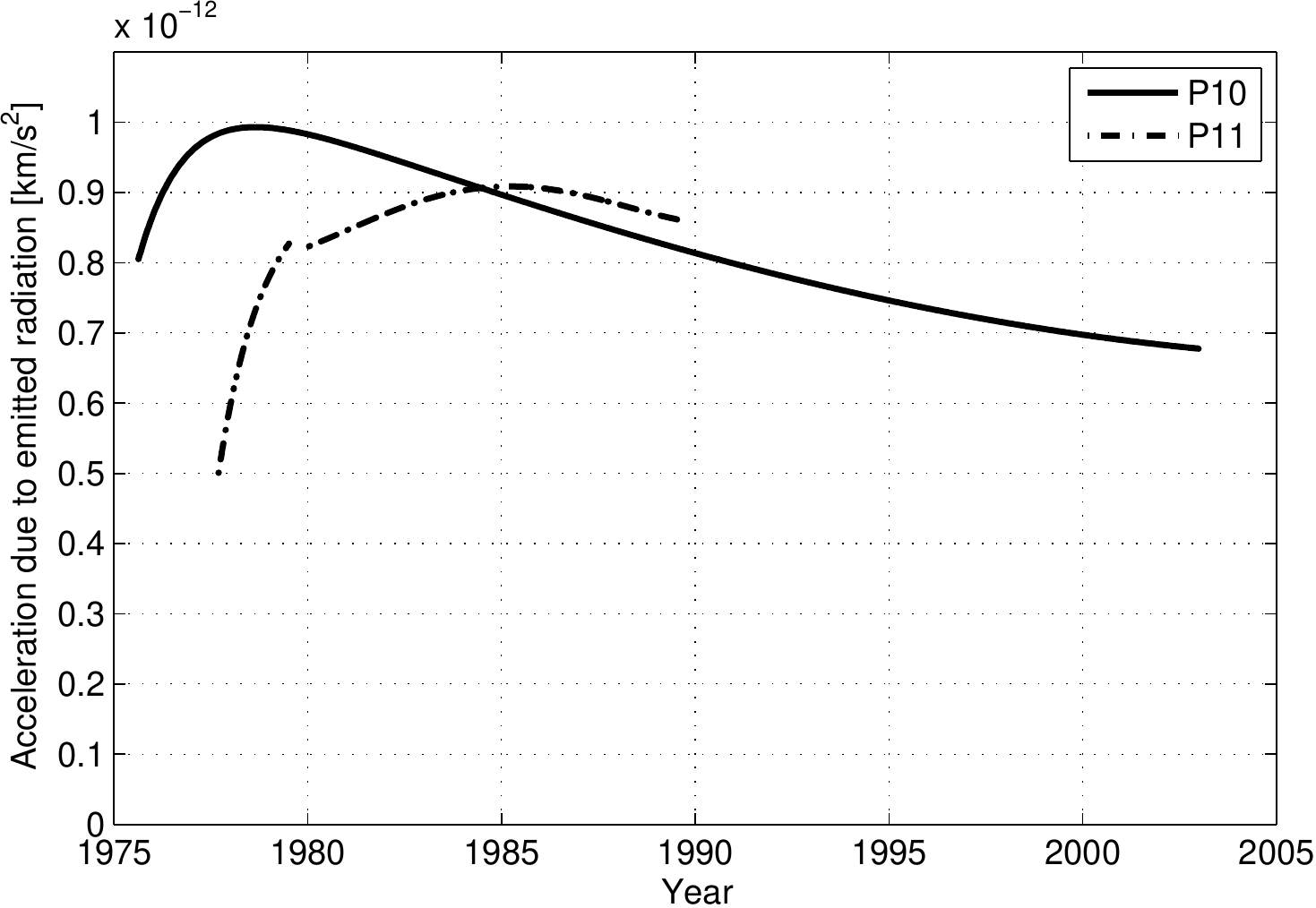}
\end{figure}

Mapping of the covariance matrix from power coefficients, $\varGamma_{{x}}$, to force coefficients, $\varGamma_{{\xi}}$, can be accomplished via an orthogonal transformation, i.e. a change of coordinates $\pmb{x} \rightarrow \pmb{\xi}(\pmb{x})$.
One last source of uncertainty is introduced when computing the thermal recoil acceleration starting from the corresponding force, since the mass of the spacecraft is not exactly known. Again, an additional covariance can be computed using an orthogonal transformation, so that the total covariance of the acceleration coefficients, $\varGamma_{RA}$, can be found as: 

\begin{equation}
%\label{eq3}
\varGamma_{RA} = \frac{1}{m_0^2} \varGamma_{RF} + \frac{ \sigma_m^2}{m_0^4} \pmb{\xi} \pmb{\xi}^T
\end{equation}

where $\sigma_m$  is the mass uncertainty assumed equal to 9 kg and the nominal mass $m_0$ was set equal to 246.4 and 235.9 Kg for Pioneer 10 and Pioneer 11, respectively (see discussion in section \ref{sec:Orbital}).
The graphical representation of the curves in Eq. \eqref{RadAcc} are shown in FIG. \ref{fig:TRAfig} for both Pioneers 10 and 11; as an example, the numerical coefficients, together with the mapped 1-sigma uncertainties for Pioneer 10 radiation recoil acceleration are collected in Table \ref{tab:ATAR_P10}. 

The curves in FIG. \ref{fig:TRAfig} show clearly that the radiation acceleration undergoes a significant temporal variation: indeed, by looking from 
left to right we first see that TRA increases due to the vanishing (outward) contribution of the solar flux. Later, at sufficiently high heliocentric distances, the (sunward) contributions from the RTG’s heat reflected by the antenna backside and the heat rejected by the louver system becomes dominant and exhibit the expected decrease due to radioactive decay of the nuclear fuel.
We point out that the maximum recoil acceleration occurs for Pioneer 10 at $\simeq 16AU$, while for Pioneer 11 at $\simeq 19AU$. The maximum value acting on Pioneer 11 is lower than the one of its predecessor, since the former resided longer within the Solar System due to a second planetary encounter at Saturn. Hence, Pioneer 11 reached distances where the effect of solar flux was negligible later in its operational life, when the available on board power had already significantly decreased.

\begin{table}%[H] add [H] placement to break table across pages
 \caption{Fitting coefficients of Pioneer 10 recoil acceleration as a function of time.}
 \label{tab:ATAR_P10}
 %\small
 \begin{ruledtabular}
 \begin{tabular}{ l c  c  }
  %\multicolumn{2}{c}{Coefficients} & Estimated value &   1-sigma \\
  {Coefficients} & Estimated value &   1-sigma \\
  \hline
 % \hline 
 $A_{RA}$		$ [km/s^2]$	&	$1.246 \times 10^{-12}$ 	&	$5.49  \times 10^{-14}$ \\ 
 $B_{RA}$		$  [km/s^3]$	&	$-9.911 \times 10^{-22}$ 	&	$4.22  \times 10^{-23}$ \\
 $C_{RA}$		$  [km/s^4]$	&	$4.191 \times 10^{-31}$ 	&	$1.81  \times 10^{-32}$ \\
 $G_{1RA}$		$  [km/s^2]$	&	$-2.685 \times 10^{-13}$ 	&	$1.62  \times 10^{-14}$ \\
 $G_{2RA}$		$  [km/s^2]$	&	$-7.148 \times 10^{-13}$ 	&	$4.31  \times 10^{-15}$ \\
 $\beta_{1}$		$  [1/s]$	&	$-7.148 \times 10^{-13}$ 	&	\textendash \footnote {No uncertainties are given to the exponential frequency factors since they represent the mapping of the $1/r^2$ term of the solar flux, which was assumed to be unaffected by errors.} \\ 
 $\beta_{2}$		$  [1/s]$	&	$-7.148 \times 10^{-13}$ 	&	\textendash \\
 \end{tabular}
\end{ruledtabular}
\end{table}%[H] add [H] placement to break table across pages

\section{Tracking Data Analyses}
\label{sec:Orbital}
Two sets of Doppler tracking data have been analyzed during this study, which cover the time intervals from February 13\textsuperscript{th}, 1980 to March 2\textsuperscript{nd}, 2002 for Pioneer 10 and from November 1\textsuperscript{st}, 1977 to September 30\textsuperscript{th}, 1990 for Pioneer 11, using NASA-JPL's Orbit Determination Program (ODP). This code includes a model of the Solar System dynamics to compute spacecraft trajectories. Based on the computed trajectory, it further calculates the predicted radio tracking observables (the so called computed observables) between the ground stations and the spacecraft. The difference between the observed and the computed observables (the so called residuals) is then fed to a recursive filter in order to improve the knowledge of a certain set of parameters affecting the spacecraft dynamics (estimated parameters). Such parameters include, as a minimum, the spacecraft state vector at a certain epoch; other parameters may be the mass and gravity field of celestial bodies, their orbits and orientation parameters, or other quantities of interest for navigation and science. Previous orbital solutions for Pioneer 10 and 11 required the addition of a sunward acceleration of unknown origin to those computed using the implemented dynamical models, both of gravitational and non-gravitational origin, in order to obtain zero-mean residuals. Such additional acceleration has become known as the Pioneer Anomaly and lacking its inclusion the Doppler residuals show an almost constant drift of 0.4 $Hz/year$, corresponding to an unmodeled acceleration of $\approx 8.5 \times 10^{-13} km/s^2$.
The trajectory reconstruction is performed in the ODP by integrating the equations of motion, expressed in terms of the total acceleration acting on the spacecraft. The gravitational forces include central body and secondary bodies Keplerian’s point mass accelerations, higher order gravity harmonics and relativistic effects. Of course not all of these contributions are always relevant for the trajectory under study. 

\subsection{Non-gravitational accelerations}
\begin{figure}
  \caption{Comparison between the magnitude of the RA (solid) and the SRP (dashed) for Pioneer 10 (top panel) and 11 (bottom panel).}
  \label{fig:SRPvsTRA_P10&P11}
    \includegraphics[width=0.75\textwidth]{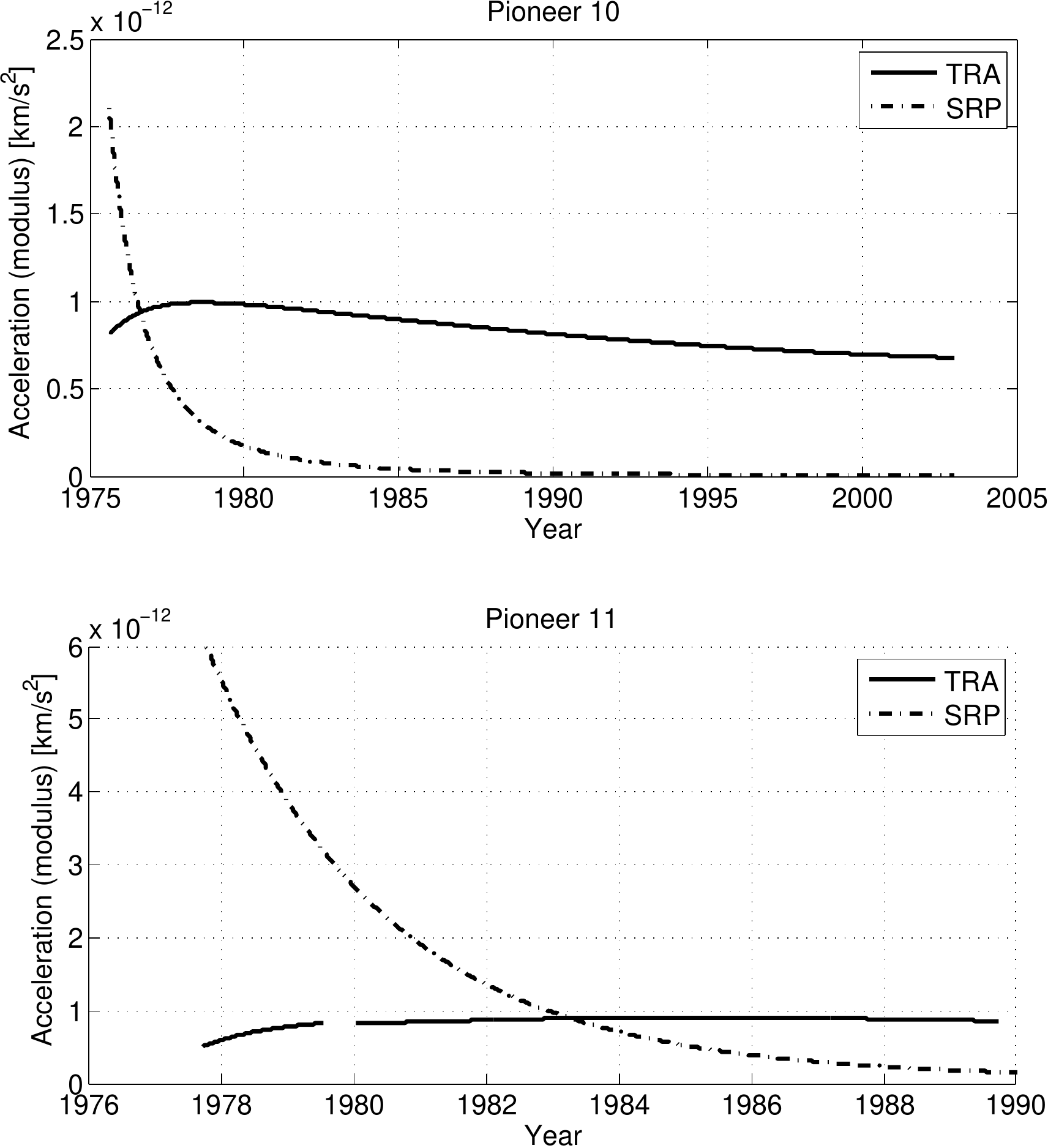}
\end{figure}

Relevant non-gravitational accelerations during Pioneers’ interplanetary cruise arise from the solar radiation pressure and propulsive maneuvers, plus, as shown in section \ref{sec:Orbital}, the acceleration due to radiation non-isotropically emitted.
The former is included as a dynamical model in the ODP in which the spacecraft parts are represented by a series of geometric entities (parabolic antenna, boxes, flat plates, spheres and cylinders). The momentum exchange between the photons and each component is computed as a function of its specular and diffuse reflection coefficients, and summed up. Because of the geometrical configuration of the Pioneers, having a big antenna dish constantly directed towards the Earth, the only component significantly contributing to the solar pressure is the antenna itself which is almost always in a full front illumination condition \footnote{at the heliocentric distances of interest for the analyzed trajectory arcs, the Earth-pointing and Sun-pointing directions are almost coincident as seen by the probes}. Using a simplified flat plate model, one can compute the following expression for the solar radiation pressure ~\cite{ODPhelp1}:

\begin{equation}
\label{SRP}
a_{SRP}= \frac{1+2(\mu_F + \nu_F)\cos(\vartheta)}{c \cdot m} \frac{A \Phi_{S@1}}{d^2}
\end{equation}

where $\mu_F$ and $\nu_F$ are the specular and diffuse reflective coefficients of HGA Earth facing side, which are assumed constant (i.e. the degradation factors have been neglected), $A$ is its area, $m$ is spacecraft mass, $\Phi_{S@1}$ is solar flux at 1 AU and $\vartheta$ is the angle between the direction of the Sun and the HGA axis. Nominal values for $\mu_F$ and $\nu_F$ coming from JPL calibration at early stages in the mission, are $8.055 \times 10^{-2}$, $2.757 \times 10^{-1}$ for Pioneer 10 and $7.016 \times 10^{-2}$, $2.808 \times10^{-1}$ for Pioneer 11. However, as reported in ~\cite{Anderson2002}, determination of these coefficients from the solar acceleration inferred from tracking data may be affected by errors in the spacecraft mass, which is not exactly known.
Equation \eqref{SRP} can be exploited to compare the relative magnitudes of the solar radiation pressure and the thermal recoil acceleration along the spacecraft trajectories. In FIG. ~\ref{fig:SRPvsTRA_P10&P11} these quantities are plotted for Pioneer 10 and 11 for the time periods covered by the available tracking data. The relative magnitudes are quite different for the two spacecraft: Pioneer 10 has a solar radiation pressure acceleration which is almost always smaller than the thermal recoil, even negligible for a large part of the trajectory. On the contrary, during Pioneer 11's route from Jupiter to Saturn, solar radiation pressure is roughly one order of magnitude higher than thermal recoil, while in the later stages of the cruise the two non-gravitational accelerations are comparable. This different dynamics has an impact on the observability of the anomalous acceleration using tracking data, as it will be discussed later in Section \ref{sec:Re-estimating}.

% NEW SECTION! 
As far as propulsive forces are concerned, the hydrazine thrusters on board the spacecraft were aimed at three types of maneuvers: precession maneuvers, i.e. HGA re-pointing towards the Earth to guarantee communication link, delta-V maneuvers for trajectory control, and spin/de-spin maneuvers. After the planetary encounters, only precession maneuvers were performed (more than one hundred in the time period covered by the analyzed data sets). Even if the precession maneuvers are expected to exert only torques on the spacecraft and no net forces, possible thrusters malfunctioning, due for example to asynchronous operation of thrusters or valve leaks, may have given rise to small residual forces. The times at which maneuvers were executed  are available through telemetry, together with the records of the commanded thrusters pulses. From this data, however, it is not possible to infer the magnitude of the (unintended) velocity increments possibly produced during maneuvers. Rather, the only means of estimating such velocity increments is using the radiometric data, i.e. to treat them as parameters to be estimated in the orbit determination analysis and check if such parameters are actually observable and/or they improve the overall quality of the fitting. This was the case for all the precession maneuvers analyzed in this study.

Non-gravitational accelerations are mass dependent, therefore the spacecraft mass must be provided as an input. Pioneers masses were nearly 259 kg (223 kg of dry mass plus 36 kg of propellant) at launch and this value was expected to decrease along the course of the mission because of propellant consumption. However, the mass was not telemetered and its value after the planetary encounters could only be reconstructed approximately. In ~\cite{Anderson2002} and ~\cite{Turyshev2010} reference values for Pioneer 10 mass are 241 and 251.8 kg, while for Pioneer 11 the reported figures are 232 and 239.7 kg. In the present study the average values of 246.4 and 235.9 kg were used as nominal masses for Pioneer 10 and 11, respectively. The uncertainty associated with these values was set to 9 kg, around one quarter of the propellant mass ~\cite{Anderson2002}, and was accounted for in the computation of the TRA as an additional covariance for the polynomial coefficients (see discussion in Section \ref{sec:MC}). 

\subsection{Media calibrations}
The ODP implements accurate models to account for media and antenna corrections to the propagation of tracking signals. Media corrections consists of corrections due to the Earth's troposphere and corrections due to charged particles which can be in the Earth's ionosphere, in space (interplanetary plasma) or in the solar corona ~\cite{Moyer2000}. The delay due to the solar corona is computed by a built-in ODP model (see ~\cite{Moyer2000} and references therein), while other effects can be accounted for if the user provides as input the zenith path delay in form of polynomials or Fourier series coefficients. In this study, tropospheric effects have been included in the form of seasonal corrections for the dry and wet delay, while corrections for ionosphere path delays based on Klobuchar's work ~\cite{Klobuch1975} were initially included but then disregarded since they did not provide a substantial improvement to the orbital fit. The ODP further includes the possibility of correcting the computed residuals for any other possible phenomenon affecting them. An example directly applicable to Pioneers is the calibration of the to account for the Doppler shift induced by the spacecraft spin rate (the so called Marini's effect).

\subsection{Implementation techniques for the orbital analyses}
\label{sec:SAvsMA}
% NEW SECTION
The data analysis using the ODP was aimed at obtaining satisfactory orbital solutions incorporating the recoil acceleration, possibly without adding any other unknown acceleration. Even if the computed recoil accelerations shown in FIG. ~\ref{fig:TRAfig} are time varying in contrast with the acceleration reported as constant in ~\cite{Anderson2002}, or decreasing monotonically ~\cite{Turyshev2010}, one should keep in mind that the measured unknown acceleration is actually a Doppler shift in the radiometric data, while the reported solutions for accelerations are just one way to obtain good orbital solution (i.e. a satisfactory fit of tracking data). Other orbital solutions may be investigated, based on dynamical models which differ from one single constant acceleration, and possibly relying on a physical basis. Indeed, the thermal analysis presented in Section \ref{sec:thermal model} provides one such model.

The implementation of the orbital analyses presented relied on two different filtering techniques, which are discussed in the following. 

In principle the trajectory followed by a spacecraft, independently from its time length, can always be fitted by a single orbital arc function of the initial spacecraft state vector, plus of every other parameter which affects its dynamics. It is therefore reasonable to include all tracking data available for a certain spacecraft into a single-arc analysis, so that all the observables contribute to the orbit determination. On the other hand, the complexity of the physics underlying certain spacecraft dynamics, especially in the presence of extremely long arcs, makes it highly improbable that the trajectory can be perfectly represented by a single deterministic model, thus, in practice, one must deal with a certain degree of model deficiency.

Pioneers tracking data, lasting more than a decade, are likely to be prone to such a problem. To overcome these difficulties, one may exploit the use of a dynamic compensation through multi-arc filtering, which has been widely used in Cassini spacecraft's scientific investigation ~\cite{IessEtAl2007}. With this method, orbital fits are obtained from shorter data arcs (from 6 months to 1 year in the present study). In the multi-arc technique the set of estimated parameters is separated into two groups: global parameters, common to all arcs, and local ones which affect only the arc to which they belong to. 

For Pioneers the parameters which may be treated as local, other than initial state vectors, are the maneuvers velocity increments; the ``anomalous" acceleration is set as global parameter so that all the available tracking data concur to its estimate.
For each trajectory arc, the orbit determination steps are performed independently, up to the computation of the observation residuals and their partial derivatives with respect to the local and global parameters. These are then combined to allow their processing by the estimation filter.
The start and end times for each arc were set at maneuver occurrences; this way, there is a certain similarity with the single-arc approach, as both allow for trajectory discontinuities at maneuvers: the former allows for velocity instantaneous increment, while the multi-arc allows for both velocity and position increments. 

The a-priori state vector components at the beginning of each arc were generated by mapping the single-arc orbital solution to the epochs of interest. Their a priori uncertainties were set equal to ten times the a-posteriori uncertainty from the single-arc solution for position vector; on the contrary, the uncertainty of the velocity components were kept completely unconstrained to allow for correct maneuver estimate.
For the single-arc analysis, the parameters to be estimated are the initial state vector components, the velocity increments due to maneuvers and a constant acceleration, consistently with the multi-arc approach. From an implementation point of view, the standard single arc filter is just a special case of the more general multi-arc filter.

Other parameters were added as consider, and their uncertainty accounted for in the computation of the formal error of the estimated parameters (see Section \ref{tab:consider}): these include the HGA reflective coefficients for SRP computation, the solar corona parameters, the tropospheric zenith path delay and the Earth stations locations. 

\begin{table} %add [H] placement to break table across pages
 \caption{Parameters treated as consider with their a-priori uncertainty (the nomenclature follows the ODP syntax, see footnote for explanations).}
 \label{tab:consider}
 \begin{small}
% \begin{ruledtabular}
 \begin{tabular} { c c }
 \hline
 \hline
  Parameter	&	A-priori Sigma \\
  \hline
%  \hline
  MUF\footnote{MUF, NUF = $\mu_F$, $\nu_F$ coefficients of HGA Earth pointing side; CORONA, CORONB, CORONC = characteristics constant for the solar corona path delay model; TROPODj, TROPOWj = constant bias to the tropospheric zenith dry and wet path delays at DSN Complex $j$; LOii, CVii, CUii = Longitude, height above equator and spin axis distance of Earth Station $ii$.}  		& $0.015$	\\
  NUF		& $0.056$		\\
  CORONA	& $4 \times 10^3m$	\\
  CORONB	& $1.8 \times 10^2m$	\\
  CORONC	& $0.8 \times 10^6m$	\\
  TROPDj 	& $5 \times 10^{-2}m$	\\
  TROPWj	& $1 \times 10^{-1}m$	\\
  LOii		& $1.57 \times 10^{-5} deg$	\\
  CVii		& $1 \times 10^{-1}m$	\\
  CUii		& $1 \times 10^{-1}m$	\\
 \hline
 \hline
 
  \end{tabular}
%\end{ruledtabular}
\end{small}
\end{table}%[H] add [H] placement to break table across pages
 
%\subsection{Data Editing}
%\label{DataEdit}
Doppler data were edited including spin compensation, data rejection and data weighting. Spin compensation was carried out according to \footnote{The total spin-induced shift accounting for the phase cycles added to the up-link and down-link signals is $(1+\alpha_{S/S})f_{SPIN}$, where $f_{SPIN}$ is the spin frequency of the spacecraft and $\alpha_{S/S}$ is the transponder turnaround ratio, which for an S/S band uplink/downlink configuration is equal to $240/221$}. For the present analyses, data were rejected when tracking from an elevation angle lower than 20 deg. Furthermore, clear outliers and biased points were manually detected and deleted. 
Doppler observables were weighted in the least square estimation filter according to the standard deviation of their residuals computed on homogeneous sets of data. To this aim, an automatic routine for data weight assignment was implemented: this allowed the post-fit sum of squares to be slightly lower than the number of observables, thus avoiding data over-weighting.

\section{Results}
\label{sec:Results}
\subsection{Re-estimating the unknown acceleration}
\label{sec:Re-estimating}
The first analysis performed consisted in re-estimating a constant acceleration using the extended data set of Pioneer 10 and 11, without accounting for the output of the thermal model discussed in Section \ref{sec:thermal model}, using both the single-arc and the multi-arc approaches (see Table \ref{tab:result1}). The formal uncertainties of the estimated accelerations are reported for all test cases, along with the corresponding values when including the consider parameters written between round brackets. This first test group is identified with the number 1, followed by an indicator of the spacecraft (P10 or P11) and the filter (SA, MA). For Pioneer 10 both single-arc and multi-arc techniques were used (first two rows in Table \ref{tab:result1}). For the MA a total of 46 arcs were implemented, each bounded in between two maneuvers and lasting approximately 6 months. 
For Pioneer 11 there are two well separated trajectory segments, the first encompassing the Jupiter to Saturn transfer orbit and a longer data set for the post-Saturn encounter trajectory. These have been treated as single arcs (rows 3 and 4 in Table \ref{tab:result1}), as the combination of two arcs (labeled with 2A to indicate a multi-arc with only 2 arcs), but also as a set of multiple arcs. Indeed, due to the very large number of maneuvers, it would have been unpractical to create one arc between each couple of maneuvers; moreover, it should be noted that the MA dynamic compensation has been motivated mainly by the evidence of periodic signatures of half a year in the residuals ~\cite{Courty2009} and annual oscillatory term in the acceleration \cite{Anderson2002}, which indicate the time scale at which modeling errors become significant. We thus used for Pioneer 11 a total of eleven arcs lasting approximately one year, each of them having a number of maneuvers treated as local parameters.

\begin{figure}
  \caption{Post-fit residuals obtained for Pioneer 10 after estimating for a constant acceleration, corresponding to test case 1.P10/MA of Table ~\ref{tab:result1}.} \label{fig:0_P10}
    \includegraphics[angle=270, width=0.75\textwidth]{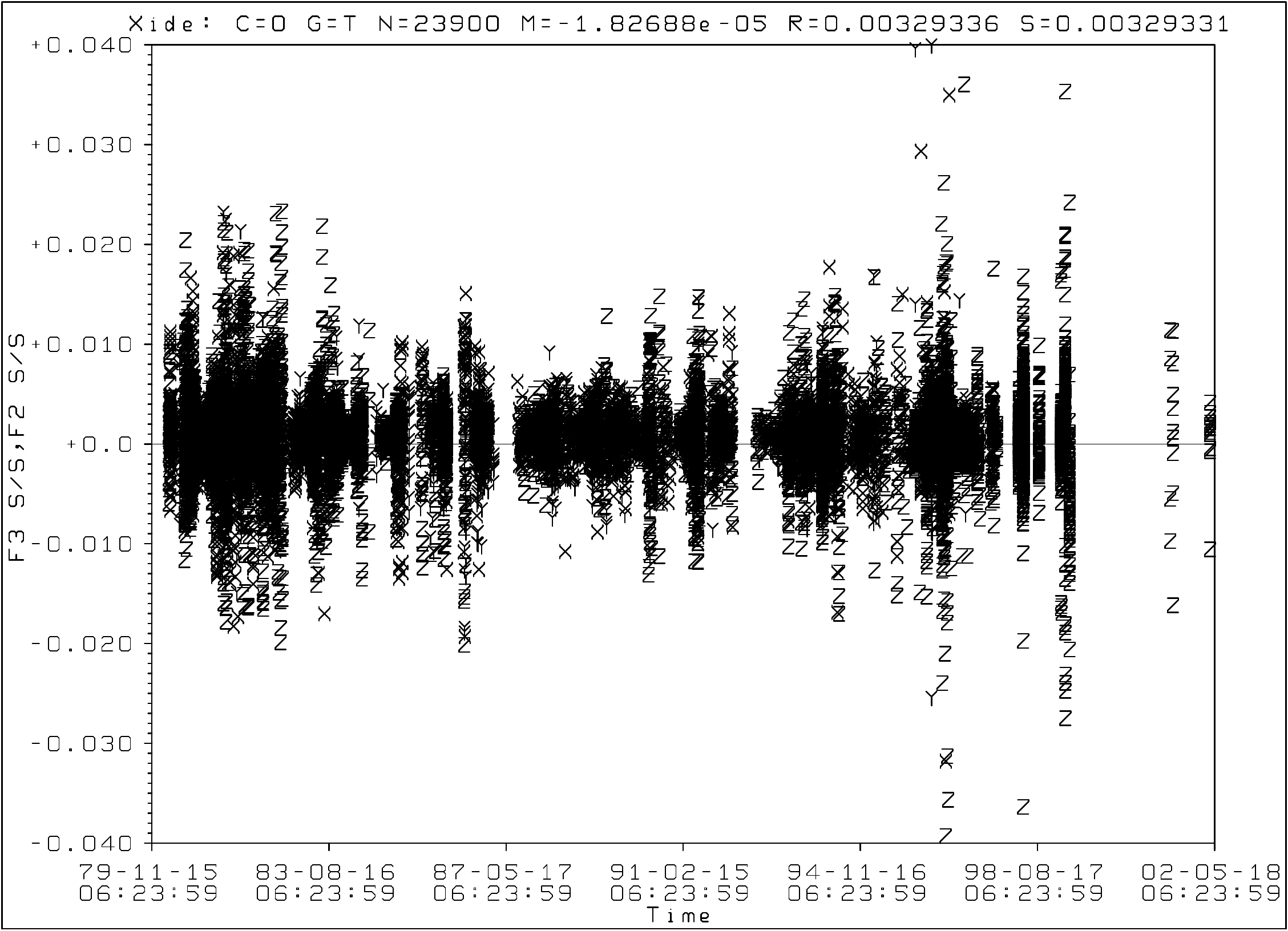}
\end{figure}

\begin{figure}
  \caption{Post-fit residuals obtained for Pioneer 11 after estimating for a constant acceleration, corresponding to test case 1.P11/2A of Table ~\ref{tab:result1}.}  \label{fig:0_P11}
    \includegraphics[angle=270, width=0.75\textwidth]{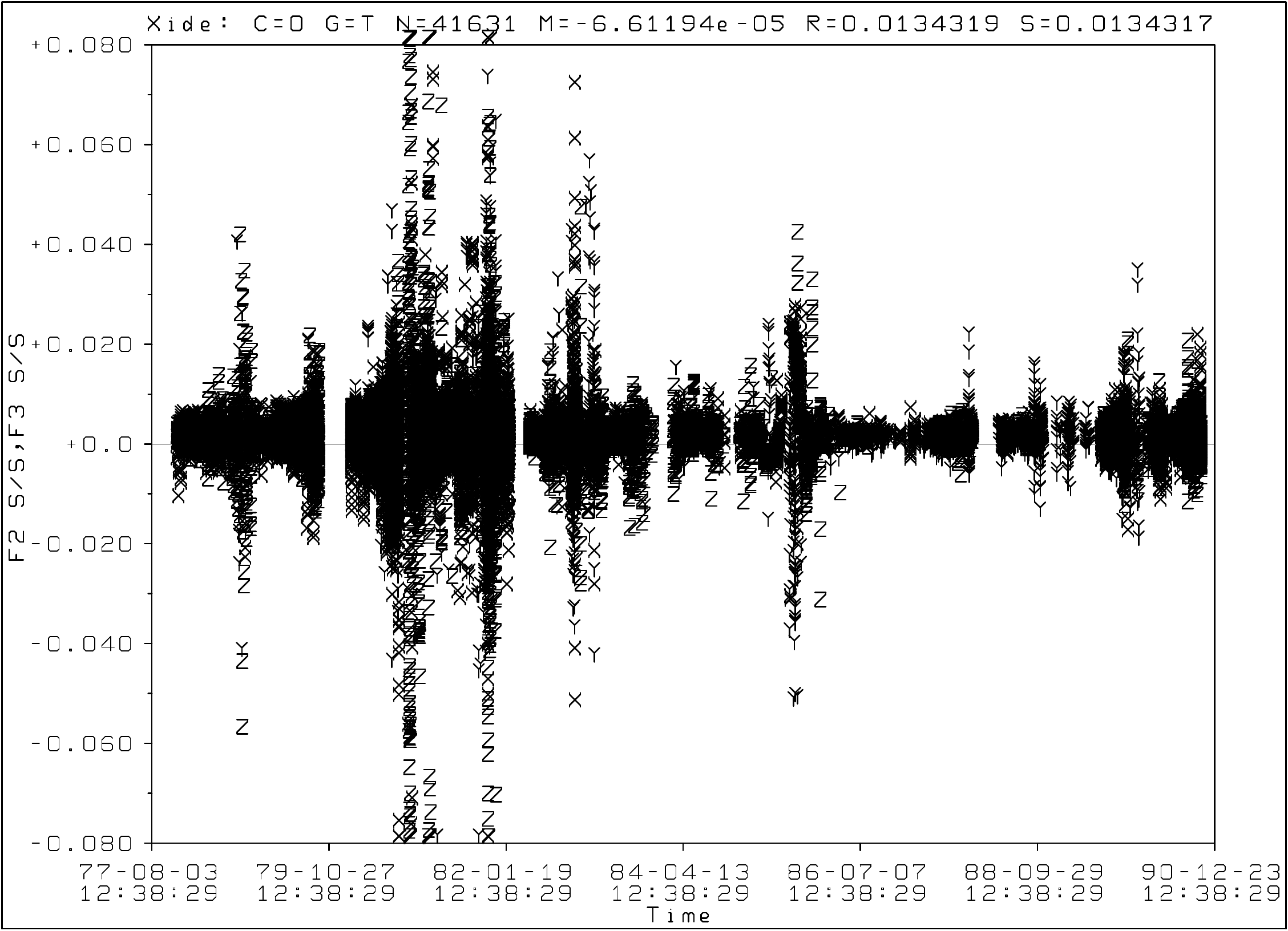}
\end{figure}

The different orbital solutions obtained are compared in terms of their mean ($\mu$) and standard deviation ($\sigma$) of the residuals and shown in Table ~\ref{tab:result1}. 
As a general trend, satisfactory orbital solutions are obtained in all cases, with some advantages in terms of lower residuals' standard deviation when using over-parameterization through the multi-arc approach. As an example, plots of Doppler residuals for the whole time span covered by the available tracking data are shown in FIG. ~\ref{fig:0_P10} and ~\ref{fig:0_P11} for Pioneer 10 and 11, respectively. 

\begin{table}%[H] add [H] placement to break table across pages
 \caption{Summary of Test Set 1 with estimation of a constant acceleration, TRA not included. Acceleration uncertainties including the consider parameters are reported in round brackets.}
 \label{tab:result1}
 \begin{ruledtabular}
 \begin{tabular}{ c cc c }
  Test case & \multicolumn{2}{c}{Post-fit residuals}  & Acceleration \\
            &   $\mu$ [mHz] & $\sigma$ [mHz]  &     $[km/s^2] \times 10^{-13}$ \\
  \hline
 1.P10/SA   &   -0.01& 3.8 & 8.18 $\pm$ 0.01 (0.08) \\
 1.P10/MA   &   -0.02& 3.3  &  7.90 $\pm$ 0.05 (0.06) \\
 1.P11-preS \footnote{prior to the Saturn encounter.} /SA &   -0.10&  4.3  & $0.89 \pm 0.60 (4.40)$\\
 1.P11-postS \footnote{after the Saturn encounter.} /SA &   -0.06&  14.1  & $7.61 \pm 0.16 (0.52)$\\
 1.P11/2A &   -0.07&  13.4  & $7.06 \pm 0.15 (0.76)$ \\
 1.P11/MA & -0.06& 13.4& $7.15 \pm 0.19 (0.82)$ \\

 \end{tabular}
\end{ruledtabular}
\end{table}%[H] add [H] placement to break table across pages
 
The estimated acceleration values obtained from these preliminary test cases are quite similar to those of other references. It should be noticed that Pioneer 10 post-fit residuals exhibits lower standard deviation levels than Pioneer 11 counterparts.
As for the pre-Saturn encounter data of Pioneer 11, the value of the acceleration is statistically null, in agreement with what reported in ~\cite{Turyshev2011}. This result is no surprising as in this part of the trajectory the solar radiation pressure is dominant with respect to the acceleration of thermal origin (see bottom panel of FIG. ~\ref{fig:SRPvsTRA_P10&P11}). %Moreover, a trajectory maneuver was performed on July 1978 of several meters per second which may hide the $\bigtriangleup V$ caused by any small acceleration. 
In general, tracking data of Pioneer 11 cover heliocentric distances at which the solar pressure is larger than, or at least comparable to, the thermal recoil acceleration. Indeed, if we add  $\mu_F$ and $\nu_F$ as solve-for parameters, the resulting acceleration varies considerably.
This is an indication that the actual magnitude of the anomalous acceleration is correlated with the solar pressure, or equivalently, a portion of the anomalous acceleration may be due to errors in modeling of solar radiation pressure. The first segment of Pioneer 11 trajectory is subjected to this to an even higher extent: by looking at the acceleration uncertainty of test case 1.P11-preS/SA it is clear how an acceleration of order of magnitude $10^{-12} km/s^2$ or less is hardly, if not at all, observable. 
Furthermore, it can be noticed that the over-parametrization has a beneficial effect on the quality of the orbital solution of Pioneer 10, when looking at the residuals standard deviation which lowers from 3.7 to 3.3 mHz.
The reduction of residuals standard deviation comes at the expense of a loss of formal accuracy (increase in the consider sigma) of the estimated acceleration; this is a well known effect in estimation filters when the number of solve-for parameters is increased without injecting additional information from other observations.

\begin{figure}
  \caption{Standard deviation of Pioneer 10 post-fit residuals calculated over 15-days batches (full line) and SEP angle (dotted line) variation from 1980 to 2000.}   
  \label{fig:Std_SEP_fig}
    \includegraphics[width=0.75\textwidth]{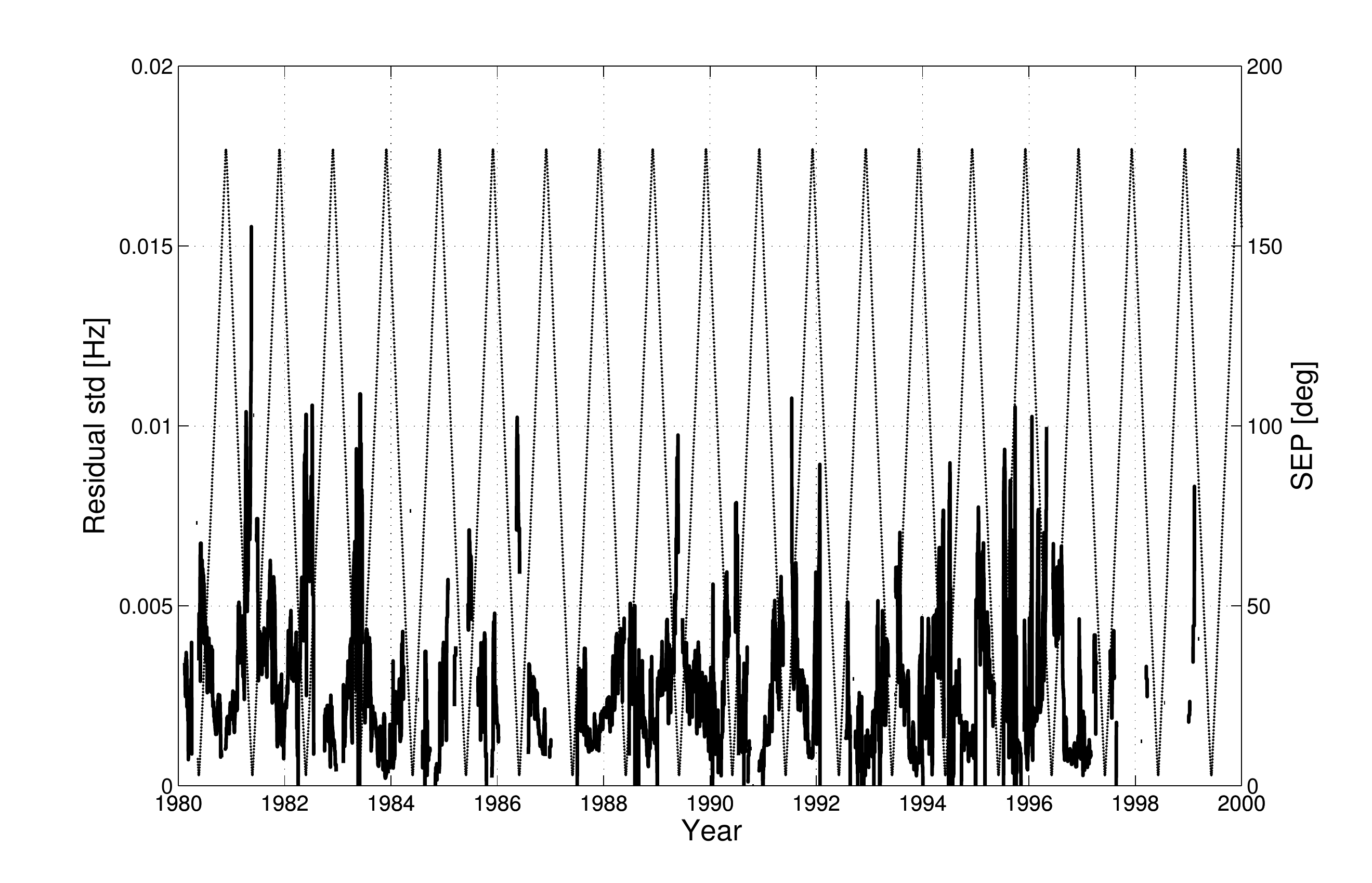}
\end{figure}

As mentioned in section \ref{sec:SAvsMA}, previous studies highlighted the presence of periodic signatures in post-fit residuals, and an annual modulation of the anomalous acceleration when estimated as a stochastic process. These conclusions were drawn from analysis of the early data set made available for Pioneer 10 (1987-1998). In the present work, the periodic modulation of the anomaly has been addressed as well, this time over the entire extended data set of Pioneer 10, but at a mainly qualitative level. 
Since S-band Doppler data are known to be highly sensitive to dispersive noise sources, a likely cause of such periodic signatures is the uncompensated delay due to the charged particles found in the solar plasma and Earth ionosphere. As previously mentioned, the ODP includes a model for solar plasma compensation, however, while such kind of model is expected to perform satisfactorily when applied to range data, on contrary it performs quite poorly for Doppler data \footnote{An empirical support to this argument comes from the fact that the orbital solutions remain practically unchanged whether the solar plasma model is included or not.}. We tested the possible correlation between the goodness of residuals and the solar plasma by comparing the temporal evolution of the standard deviation of post-fit residuals batches lasting 15 days, along with the Sun-Earth-Probe (SEP) angle. The outcome is shown in FIG. ~\ref{fig:Std_SEP_fig} where several standard deviation peaks are found in correspondence of low SEP angles.
Starting from this evidence, it is quite natural that any spectral analysis of the residuals may highlight annual peaks; at the same time, given the periodic modulation of the Doppler signal, an acceleration with the same frequency can compensate for it, improving the data fit: the previously reported results \cite{Anderson2002}, in this respect, are hence of no surprise.
Summarizing, the authors are firmly convinced that the temporal modulation of the anomaly is an artifact due to imperfect media calibration, in particular solar plasma: as a consequence, this issue has not been the object of further investigation.

\subsection{Orbital solutions including the recoil force}
The preliminary orbit analyses summarized in Table ~\ref{tab:result1} have been then repeated with the following variants:
\begin{itemize}
 \item Test Set 2, where the TRA model is included in the trajectory reconstruction and treated as consider.
 \item Test Set 3, where TRA is included as consider and an additional bias acceleration is estimated (global parameter in case of multi-arc). This is an original contribution of the present study, which has not been reported in \cite{Turyshev2012}.
\end{itemize}

\begin{figure}
  \centering
  \caption{Post-fit residuals obtained for Pioneer 10, after including the recoil force dynamical model, corresponding to test case 2.P10/MA of Table ~\ref{tab:result2}.} \label{fig:1_P10}
    \includegraphics[angle=270, width=0.75\textwidth]{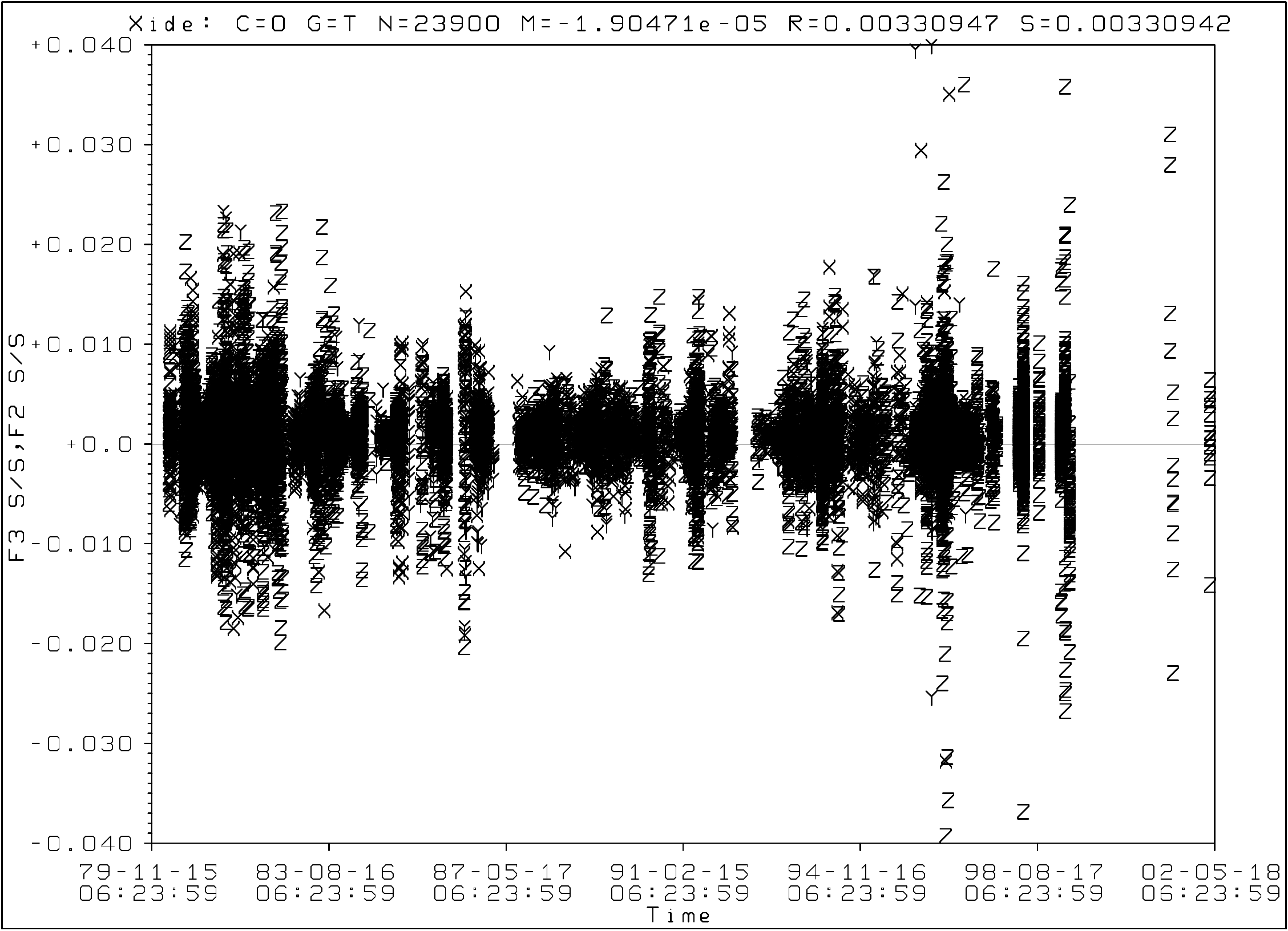}
\end{figure}

\begin{figure}
  \centering
  \caption{Post-fit residuals obtained for Pioneer 11, after including the recoil force dynamical model, corresponding to test case 2.P11/MA of Table ~\ref{tab:result2}.} \label{fig:1_P11}
    \includegraphics[angle=270, width=0.75\textwidth]{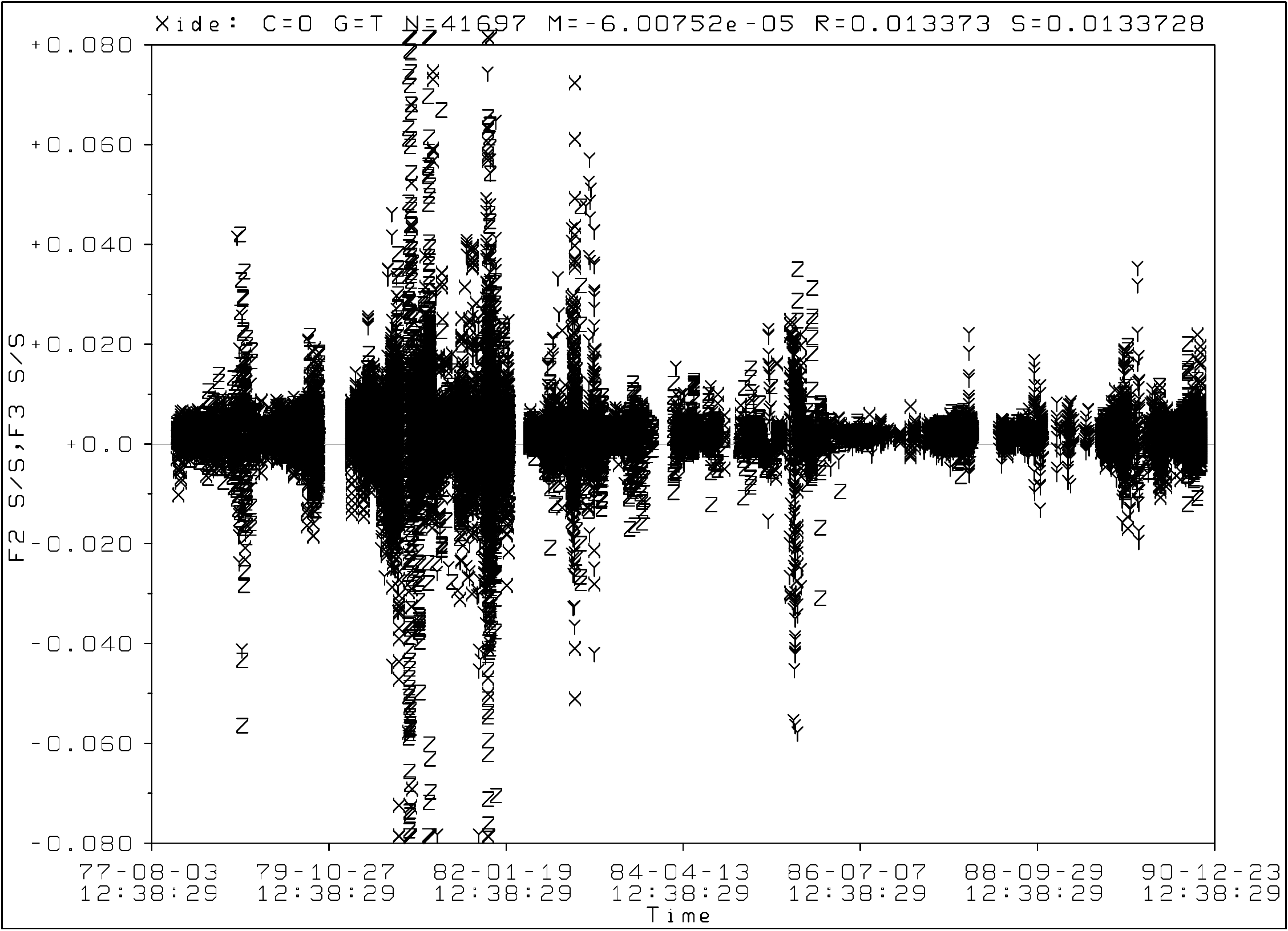}
\end{figure}

The introduction of an additional acceleration for Test Set 3 has the purpose of checking whether thermal recoil force is enough to explain the whole anomaly, or if a residual unmodeled acceleration still provides an improved orbital solution. Parameters treated as consider which include the ones found in Table \ref{tab:consider} plus the coefficients for the recoil acceleration. Results are summarized in Table \ref{tab:result2} and \ref{tab:result3} and discussed afterwards; sample plots of post-fit residuals obtained including recoil accelerations for Pioneer 10 and 11 complete data sets are shown in FIG. ~\ref{fig:1_P10} and ~\ref{fig:1_P11}. 
Since the acceleration in the pre-Saturn encounter of Pioneer 11 is unobservable, the case 3.P11-preS is not reported in Table \ref{tab:result3}.
When including time varying accelerations according to our thermal model, orbital solution are obtained for Pioneer 10 and for the two segments of Pioneer 11, without adding any other acceleration. This is an implicit confirmation that the observed drift in Doppler residuals is compatible with time varying acceleration. 
The quality of the fit is equivalent to that of the solutions obtained with Test Set 1, as emerges by comparing FIG. \ref{fig:0_P10} and ~\ref{fig:0_P11} with the corresponding \ref{fig:1_P10} and ~\ref{fig:1_P11}, as well as the post-fit residuals statistics in Table \ref{tab:result2} and \ref{tab:result3}. In particular, in the residuals displayed in FIG. \ref{fig:1_P10} and ~\ref{fig:1_P11} no drift or signatures are present which might indicate for any residual unaccounted acceleration. 
A confirmation of this is found with Test Set 3: the additional acceleration cannot be clearly estimated, since the estimated value is not sufficiently larger than its consider sigma, this being especially true for Pioneer 10 (lower bias to sigma ratios). Most important, even when the acceleration is larger than 2-sigma, there is absolutely no gain in the orbital solution quality by the introduction of this additional term. All the above considerations hold for both Pioneer 10 and 11, as well as for single-arc and multi-arc filtering.
In summary, all simulations performed lead to the same conclusion: there is no anomalous acceleration acting on Pioneer 10 and 11. The reported unexplained drift in Doppler residuals disappears when including the force due to anisotropic radiation emission into the dynamical model of the probes.
\begin{table}%[H] add [H] placement to break table across pages
 \caption{Summary of Test Set 2, including TRA, with no additional acceleration estimated.}
 \label{tab:result2}
 %\resizebox*{0.45\textwidth}{!}{
 \begin{ruledtabular}
 \begin{tabular}{c cc}
  Test case & \multicolumn{2}{c}{Post-fit residuals}  \\
            &   $\mu$ [mHz] & $\sigma$ [mHz]  \\
  \hline
 2.P10/SA   &   -0.02& 3.6  \\
 2.P10/MA   &   -0.02& 3.3 \\
 2.P11-preS/SA &   -0.12 & 4.4   \\
 2.P11-postS/SA &   -0.06&  14.1   \\
 2.P11/2A &   -0.07&  13.4  \\
 2.P11/MA &   -0.06&  13.4  \\
 \end{tabular}
\end{ruledtabular}
\end{table}%[H] add [H] placement to break table across pages

\begin{table}%[H] add [H] placement to break table across pages
 \caption{Summary of Test Set 3, including TRA and the estimation of an additional constant acceleration  (acceleration uncertainties in presence of consider parameters reported in round brackets).}
 \label{tab:result3}
 \begin{ruledtabular}
 \begin{tabular}{c cc c }
  Test case & \multicolumn{2}{c}{Post-fit residuals}  & Acceleration \\
            &   $\mu$ [mHz] & $\sigma$ [mHz]  &     $[km/s^3]$ \\
  \hline
 3.P10/SA   &   -0.03& 3.6 & $0.18 \pm 0.01 (0.44)$ \\
 3.P10/MA   &   -0.02& 3.3  &  $0.83 \pm 0.05 (0.52)$\\
 %3.P11-preS/SA &   -0.01 & 4.3  & $-7.54 \pm 0.48 (4.70)$ \\
 3.P11-postS/SA &   -0.06&  14.1  & $-1.12 \pm 0.12 (0.67)$ \\
 3.P11/2A &   -0.06&  13.4  & $-1.62 \pm 0.15 (0.87)$\\
 3.P11/MA &   -0.06&  13.4  & $-1.35 \pm 0.19 (0.92)$\\
 \end{tabular}
\end{ruledtabular}
\end{table}%[H] add [H] placement to break table across pages

\section{Conclusions}
\label{sec:Conclusions}
Orbital solutions for Pioneer 10 and 11 spacecraft using the available radiometric observables were presented. The data processing was carried out taking into account the results of a detailed thermal modeling of the spacecraft aiming at evaluating the recoil force due to the anisotropic radiation. The thermal model includes the main spacecraft components and was built using the available design documentation and the retrieved telemetry data. Monte Carlo simulations allowed to perform a sensitivity analysis on the solution, and to represent the thermal acceleration along the trajectory using a finite number of parameters and their associated uncertainties. Such representation is suitable of being incorporated in the orbit determination process in a consistent manner.

Processing of radiometric data was performed using the NASA/JPL's ODP. Both single-arc and multi-arc estimation techniques were implemented and the resulting orbital solutions thoroughly compared. The multi-arc technique provides a means to compensate for deficiencies in dynamical models when the trajectory arc is extended in time, allowing for a slightly better quality of the post-fit residuals. 
The systematic estimation of velocity increments due to each propulsion maneuver was also included as a necessary step to reach a good orbital solution.

Our results show that the computed thermal recoil acceleration, though not constant in time, is the only responsible for the observed linear drift in the Doppler data reported in previous literature. Our orbital solutions were obtained without the need for any empirical acceleration in addition to the thermal recoil one. We also tried including the estimation of an additional constant acceleration, but it did not improve the quality of the orbital fits, at the same time being its estimated value statistically compatible with zero. All these results lead to the conclusion that no anomalous acceleration acted on Pioneer 10 and 11 spacecraft along their interplanetary trajectories, once all systematic effects, and in particular the thermal recoil force, are included in the dynamical model: the Pioneers follow trajectories which are fully compatible with Newton-Einstein's laws of gravity.

\subsection{Aknowledgements}
The authors are grateful to Slava G. Turyshev for his survey on Pioneer investigation status, Victor T. Toth for his advice on telemetry data, and Ruaraidh Mackenzie for his valuable hints on multiarc analysis.

The work of D. M. and P. T. has been funded in part by ASI (Agenzia Spaziale Italiana). 

\bibliography{PioneerBib}

%\rightarrow
\end{document}